# Atomic structure and electronic properties of nanotubes of layered iron-based superconductors.

I. R. Shein, A. N. Enyashin, and A. L. Ivanovskii *

*Institute of Solid State Chemistry, Ural Branch of the Russian Academy of Sciences, 620990, Ekaterinburg, Russia*

**A B S T R A C T**

The atomic models of nanotubes for layered FeSe, LiFeAs, $SrFe_2As_2$, and LnFeAsO - the parent phases of so-called 11, 111, 122, and 1111 groups of newly discovered family of iron-based high temperature superconductors are proposed. On example of $SrFe_2As_2$ the electronic properties of predicted nanotubes are examined and discussed in comparison with those for the corresponding single layer and the crystal.

* Corresponding author.
*E-mail address:* shein@ihim.uran.ru (I.R. Shein).

# 1. Introduction

The discovery of superconductivity in single- and multi-walled carbon nanotubes (NT) as well as in their ropes (see [1-6]) has inspired the further intensive search of superconducting systems among other non-carbon nanotubes, that may open up new possibilities for the fundamental understanding of the effects of dimensionality on superconductivity.

One of the most obvious ways in this direction is to design the nanotubes of known bulk (3D) superconductors, while expecting their one-dimensional (1D) nanoscale forms also preserve the superconductivity. Clear and successful evidences of this approach have been obtained on example of $MgB_2$, which in the bulk state adopts superconductivity near $T_c \sim$ 40K [7]. Recently, single-crystalline tube-like nanostructures of $MgB_2$ were fabricated, which demonstrate superconducting properties at ~37K [8]. Note, that $MgB_2$ belongs to a rich set of layered compounds and is a highly anisotropic quasi-two-dimensional system, where metal layers alternate with graphene-like boron layers. Such layered compounds possess a significant structural flexibility and belong to the most promising materials for formation of the nanotubes and other related cage-like nanostructures, see for example [9,10].

In this context, the recent discovery [11] of a new family of high-temperature iron-based superconductors (SCs) with $T_c$ up to 56K and with unconventional pairing mechanism near to the spin-density wave order seems to be highly interesting. Indeed, all of these Fe-based SCs (quaternary, ternary Fe pnictides and binary Fe chalcogenides) have quasi-two-dimensional crystal structures and contain FeAs (or FeSe) layers as the superconducting blocks, which alternate with positively charged insulating blocks - so-called "charge reservoirs", reviews [12-17]. The available data reveal that bonding picture in these layered species has a complex and very anisotropic character. The strong covalent bonding appears inside FeAs (or FeSe) layers, whereas between the adjacent layers quite weak ionic bonds occur. Thus, these compounds (which possess the considerable structural and chemical flexibility [12-20]) should be very sensitive to the dimensionality effects. Recently, the first attempts to study this effect have been performed by a comparison of 3D (bulk) and 2D (surface) states for some Fe-based SCs [21-24].

We assume that the above mentioned layered Fe-based SCs can exist as nanotubes (NTs) with unusual conductive and magnetic properties. A first key step could be a characterization of atomic structure and electronic properties of these nanotubes.

In this work we focus on the description of the atomic models for proposed NTs of layered FeSe, LiFeAs, SrFe$_2$As$_2$, and LnFeAsO - the parent phases of so-called 11, 111, 122, and 1111 groups of iron-based superconductors. Besides, on the example of SrFe$_2$As$_2$ NTs their electronic properties are examined and compared with those for the corresponding single layer and the crystal.

## 2. Structural models

Nowadays, four main so-called 11, 111, 122, and 1111 groups of iron-based SCs have been fabricated; the parent phases for these groups are FeSe, LiFeAs, SrFe$_2$As$_2$, and LnFeAsO, respectively. All these phases have the common structural motifs [12-17].

Quaternary tetragonal arsenide-oxide LaFeAsO has a tetragonal layered structure (space group *P*4/*nmm*, Z=2), where [LaO]$^{\delta+}$ layers are sandwiched between [Fe$_2$As$_2$]$^{\delta-}$ layers; the atomic positions are La: 2*c* (¼,¼,$z_{La}$), Fe: 2*b* (¾,¼,½), As: 2*c* (¼,¼,$z_{As}$), and O: 2*a* (¾,¼,0). The ternary arsenide SrFe$_2$As$_2$ adopts also tetragonal structure (space group *I*4/*mmm*; Z=2) and is formed by a stack of alternating Sr$^{\delta+}$ atomic sheets and [Fe$_2$As$_2$]$^{\delta-}$ layers; the atomic positions are Sr: 2*a* (0,0,0), Fe: 4*d* (½,0,½), and As: 4*e* (0,0,$z_{As}$). Further, the ternary arsenide LiFeAs crystallizes in a tetragonal unit cell, space group *P*4/*nmm*. This layered structure is built up of [Fe$_2$As$_2$]$^{\delta-}$ layers alternating along the *c* axis with nominal double layers of Li atoms. The atomic positions are Li: 2*c* (¼,¼,$z_{Li}$), Fe: 2*b* (¾,¼,½), and As: 2*c* (¼,¼,$z_{As}$). Finally, binary selenide α-FeSe (tetragonal, space group *P*4/*nmm*) is composed of only the [Fe$_2$Se$_2$] layers with the atomic coordinates Fe: (¼,-¼,0), and Se: (¼,¼,$z_{Se}$). Here $z_X$ are the so-called internal coordinates governing the Fe-As(Se) distances and the distortion of the {As(Se)Fe$_4$} tetrahedra.

Thus, all of these phases include [Fe$_2$As$_2$] (or [Fe$_2$Se$_2$]) layers formed by edge-shared {FeAs$_4$} (or {FeSe$_4$}) tetrahedra, where the iron atoms are arrayed in the square lattice. These layers further are used to design structural models of FeSe, LiFeAs, SrFe$_2$As$_2$, and LnFeAsO nanotubes.

In Fig. 1 a fragment of the square-like [Fe$_2$As$_2$] (or of the isostructural [Fe$_2$Se$_2$]) layer is shown, together with primitive vectors ***a$_1$*** and ***a$_2$***. Based on these vectors one can construct the infinitely long nanotubes of the simplest Fe-based SC – binary FeSe in conventional way by rolling of [Fe$_2$Se$_2$] layers into cylinders, which is by analogy to a graphene sheet forming carbon nanotubes, see for example [25].

Like a [Fe$_2$Se$_2$] flat layer, the walls of these nanotubes can be presented as three coaxial atomic cylinders: one Fe and two (external and internal) Se cylinders: Se@Fe$_2$@Se, see Fig. 2. In general, the family of FeSe NTs may be described in terms of the primitive 2D lattice vectors ***a$_1$*** and ***a$_2$*** and two integer indices $n$ and $m$: ***B*** = $n$***a$_1$*** + $m$***a$_2$***, and classified into three groups: $n = m$ "*armchair*", $n \neq 0$, $m = 0$ "*zigzag*", and $n \neq m$ "*chiral*" nanotubes.

The atomic models of nanotubes of ternary or quaternary iron arsenides will be more complicated. First of all, these NTs will contain the additional atomic cylinders which are composed of alkaline or alkaline-earth atomic sheets (e.g., Li, Sr) and LnO layers within LiFeAs, SrFe$_2$As$_2$, and LnFeAsO phases, respectively.

For example, SrFe$_2$As$_2$ NTs contain four coaxial atomic cylinders: one Fe, two As and one Sr cylinders, and these tubes also can adopt various chirality as depicted in Fig. 1. Moreover, various types of atomic configurations of NTs may correspond to one and the same formal composition of iron arsenides. For example, for LiFeAs: (i) all of Li atoms can form outer ([FeAs]@Li) or inner (Li@[FeAs]) atomic wall of the tubes or (ii) one half of lithium atoms can form outer wall, other half - inner wall, and in this case such LiFeAs NT should be described already as: Li$_{½}$@[FeAs]@Li$_{½}$ as shown in Fig. 2. Certainly, the ratio of Li atoms (or Sr atoms for SrFe$_2$As$_2$ NTs) deposited on outer and inner walls can differ from ½. On the contrary, for the NTs of quaternary 1111 phases such as LnFeAsO, owing to covalent bonding inside [LnO] layers [12-15], only two preferable types of atomic configurations should be assumed: [LnO]@[FeAs] or [FeAs]@[LnO], see Fig. 2.

Thus, we can manifest that four main 11, 111, 122, and 1111 groups of iron-based SCs can form essentially various types of nanotubes. Simplest of them are FeSe NTs: all these tubes will have identical atomic configuration (Se@Fe$_2$@Se, *i.e.* these NTs are Se-terminated), and these tubes will differ only by their diameters and chirality. For 1111 based NTs, beside their diameters and chirality, two main atomic configurations should be assumed: arsenic-terminated [LnO]@[FeAs] or oxygen-terminated: [FeAs]@[LnO]. Finally, the largest set of possible atomic configurations should be assumed for tubes based on 111 or 122 phases, which include the sheets of atoms of alkaline or alkaline-earth metals, which are bonded with [Fe$_2$As$_2$] layers by ionic interactions. These atoms can cover the outer or/and inner walls of these NTs with different density and arrangement types. Thus, the formal configuration, for example, of LiFeAs tubes, may be described as Li$_x$@[FeAs]@Li$_{1-x}$.

## 3. Electronic properties

Let us consider in more details the structural, electronic properties and stability of proposed nanotubes - on example of SrFe$_2$As$_2$ NTs. Here our ambitions are to answer (i). Whether these NTs are stable? (ii). What are the features of their structural and electronic properties? (iii). How these properties depend on possible atomic configurations of SrFe$_2$As$_2$ NTs?

For this purpose we have chosen "*armchair*" (9,9) SrFe$_2$As$_2$ tubes with two main configurations: 1 – [FeAs]@Sr, *i.e.* Sr-terminated, and 2 - Sr@[FeAs], *i.e.* As-terminated. Additionally we have studied the bulk SrFe$_2$As$_2$ and single flat slab of this arsenide.

For the calculations of infinitely long (9,9) SrFe$_2$As$_2$ nanotubes, the unit cells containing 45 atoms have been used. Periodic boundary conditions were employed along the nanotube axis, and the vacuum region of about 1.4 nm is assumed between the tube and its images in order to exclude tube-tube interaction.

Our first-principles calculations of mentioned tubes, as well as bulk SrFe$_2$As$_2$ and its single slab have been performed using the SIESTA code [26]. The valence electrons were described by linear combination of numerical atomic-orbital basis set and the atomic core by norm-conserving pseudopotentials. The pseudopotentials generated using the Troullier and Martins scheme [27] were applied to describe the interaction of valence electrons with atomic core and their nonlocal components are expressed in the full separable form of Kleinman and Bylander [28,29]. The generalized gradient approximation (GGA) formalism [30] is adopted for the exchange-correlation potential. The Hamiltonian matrix elements are calculated by the charge-density projection on a real-space grid with an equivalent plane-wave cutoff energy of 300 Ry. We used Monkhorst–Pack $k$ points originating at the Γ point 1×1×25 for all structures. The conjugate gradient algorithm [31] was applied for the relaxation of the all examined systems until the maximum force on a single atom is within 0.02 eV/Å.

The results obtained are presented on Fig. 3 and in Table and allow us to draw the following conclusions.

First, we found that the optimized $SrFe_2As_2$ NTs retain the cylindrical-like shapes of their initial constructed atomic models, indicating the stability of these hollow tubular structures. The calculated inter-atomic distances ($d$, see Table) show that for bulk $SrFe_2As_2$ these values are in reasonable agreement with the available experimental data [32]. The deviations of $d$ for tubes from these equilibrium distances arise mainly from curvature of atomic layers and depend also from atomic configurations of the tubes and relaxation effects.

The next important point involves the relative stability of $SrFe_2As_2$ NTs with various types of atomic configurations. Our calculations of the total energies ($E_{tot}$) for examined $SrFe_2As_2$ NTs with alternative atomic configurations (see above) demonstrate that the As-terminated NTs should be much less stable comparing to Sr-terminated tubes, see Table. This fact can be easily explained by strong repulsive interactions between $Sr^{2+}$ cations inside inner cylinders for Sr@[FeAs]-like NTs, where the inter-atomic distances are significantly shorter ($d$(Sr-Sr) = 3.22 Å) - as compared with the equilibrium distances ($d$(Sr-Sr) = 3.96 Å) in the bulk $SrFe_2As_2$. On the contrary, for Sr-terminated NTs these distances became increased: $d$(Sr-Sr) ~ 5.20 Å. Thus, a formation of $SrFe_2As_2$ tubes should be accompanied by the tendency of the atoms of alkaline-earth metals to be located on the external sides of the tubes. The similar situation should be expected also for related 111-based NTs, where the atoms of alkaline metals should favor to form their outer walls.

For examined $SrFe_2As_2$ NTs we have also estimated the strain energy $E_{str}$, which are defined as the difference between the total energies $E_{tot}$ of a tube and the corresponding flat slab. For example, $E_{str}$ for more stable Sr-terminated NT (9,9) tube is close to 0.054 eV/atom, which is in between the values $E_{str}$ for massively fabricated carbon nanotubes ($E_{str}$ ~ 0.01 eV/atom [33]) and inorganic $MoS_2$ tubes ($E_{str}$ ~ 0.15 eV/atom [34]) with comparable diameters. This estimation may be considered as an additional argument for the possibility of an existence of the proposed $SrFe_2As_2$ tubular forms.

The calculated densities of states (DOSs) for $SrFe_2As_2$ bulk, slab and tubes are shown in Fig. 3. For crystalline $SrFe_2As_2$, the overall features of the DOS shape are in good agreement with previous calculations [19,35,36]. The upper valence band and the lower conduction band are dominated by the Fe $3d$ states with admixtures of the As $4p$ states; suggesting strong covalent $p$-$d$ hybridization between As and Fe. The contributions from the valence Sr states to the occupied bands are negligible, *i.e.* in $SrFe_2As_2$ these atoms are in the form of cations $Sr^{2+}$, see also [37].

For discussed iron-based superconductors the most interesting is the distribution of the electronic states in the window around the Fermi level $E_F$, which arises mainly from the Fe $3d$ bands. The total DOSs at the Fermi level, $N(E_F)$, for examined 3D-1D $SrFe_2As_2$ systems are shown in Table. It is seen that as going from 3D (bulk) to 2D state (slab), the value of the $N(E_F)$ increase at about 24 %. For $SrFe_2As_2$ nanotubes the values of the $N(E_F)$ depend strongly on their atomic configurations (see Table) but remain also higher than $N(E_F)$ for crystalline $SrFe_2As_2$. Thus, the lowering of dimensionality (3D→ 1D) for $SrFe_2As_2$ results in a significant redistribution of near-Fermi bands, which can lead to the occurrence of unusual transport and magnetic properties for such nano-scale materials.

Finally, we speculate, that the suggested tubes can likely be prepared, for example, using the Prinz's technology from thin epitaxial films, which can be controllably detached from substrate and rolled into cylindrical micro- and nanotubes, see [38,39].

## 4. Conclusions

In summary, the atomic models of nanotubes of layered FeSe, LiFeAs, $SrFe_2As_2$, and LnFeAsO - the parent phases of so-called 11, 111, 122, and 1111 groups of recently discovered family of iron-based high temperature superconductors are proposed. We find that the simplest of them will be FeSe NTs. All these tubes will have identical atomic configuration (Se@$Fe_2$@Se) and will differ only by their diameters and chirality. For 1111 based NTs, beside their diameters and chirality, two main atomic configurations should be assumed: arcenic-terminated or oxygen-terminated. A larger set of possible atomic configurations arises for tubes based on 111 or 122 phases, where the atoms of alkaline or alkaline-earth metals can cover outer or/and inner walls of these NTs with different density and arrangement types.

By means of the first-principle SIESTA code, on the example of $SrFe_2As_2$ the electronic properties of predicted nanotubes are examined and discussed in comparison with those for the corresponding single layer and the bulk crystal. According to the results obtained, the starting atomic models of $SrFe_2As_2$ NTs preserve their cylindrical morphology upon geometry optimization and their strain energies are comparable with $E_{str}$ for widely known carbon or inorganic $MoS_2$ tubes. These facts testify the stability of proposed $SrFe_2As_2$ tubular forms. The decreasing of dimensionality (3D→ 1D) for $SrFe_2As_2$ leads to the growth of near-Fermi level densities of states that in turn can lead to the occurrence of unusual transport and magnetic properties for these nano-scale materials.

**Acknowledgements**
Financial support from the RFBR (Grants 09-03-00946 and 10-03-96008-Ural) is gratefully acknowledged.

**Table.**
Selected intra-atomic distances (*d*, in Å), diameters of NTs (*D*, in Å), total densities of states at the Fermi level (N($E_F$), in states/eV· form.unit), and the relative energies (ΔE, in eV/form.unit) for bulk, slab and proposed nanotubes of $SrFe_2As_2$ in comparison with available data.

|  | NT-1* | NT-2 | slab | bulk *** |
|---|---|---|---|---|
| *d*(Sr-Sr) | 5.20 | 3.22 | 3.96 | 4.02 |
| *d*(Sr-As) | 3.42 | 3.08 | 3.27 | 3.29 (3.24-3.27 [a]) |
| *d*(As-Fe) ** | 2.47/2.30 | 2.32/2.27 | 2.32 | 2.35 (2.39 [a]) |
| *d*(Fe-Fe) | 2.76 | 3.05 | 2.80 | 2.84 (2.79-2.76 [a]) |
| $D^{outer}$ | 20.05 | 19.84 | - | - |
| $D^{inner}$ | 12.28 | 12.49 | - | - |
| ΔE | 2.56 | 3.02 | 2.29 | 0 |
| N($E_F$) | 3.84 | 2.85 | 3.06 | 2.34 (2.74 [b]; 2.29 [c]) |

*for "*armchair*" (9,9) $SrFe_2As_2$ tubes with two main configurations: NT-1: [FeAs]@Sr, and NT-2: Sr@[FeAs], see also the text;
** between Fe atom and As atoms from outer/inner As cylinders
*** available data are given in parentheses: [a] experiment, Ref. [31], [b,c] calculations, [b] FLAPW-GGA code, Ref. [19]; [c] KKR-ASA code, Ref. [33].

**FIGURES**

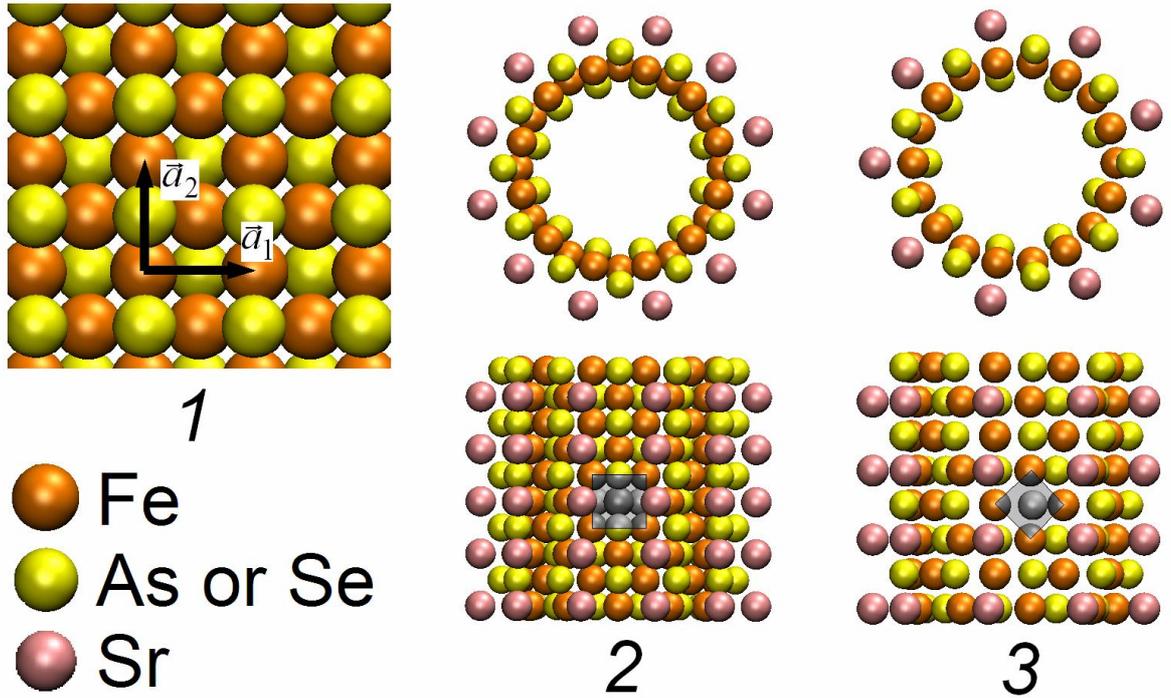

**Figure 1.** (Color online) *1*: Fragment of basic [Fe$_2$As$_2$] (or [Fe$_2$Se$_2$]) layers of 11, 111, 122, and 1111 groups for iron-based superconductors formed by edge-shared {FeAs$_4$} (or {FeSe$_4$}) tetrahedra; the primitive vectors *a$_1$* and *a$_2$* used for design of the corresponding NTs are depicted. *2* and *3* - atomic structures of *zigzag* (12,0) and *armchair* (9,9) SrFe$_2$As$_2$ NTs with configuration [FeAs]@Sr (see text), respectively. Top view and side view are shown.

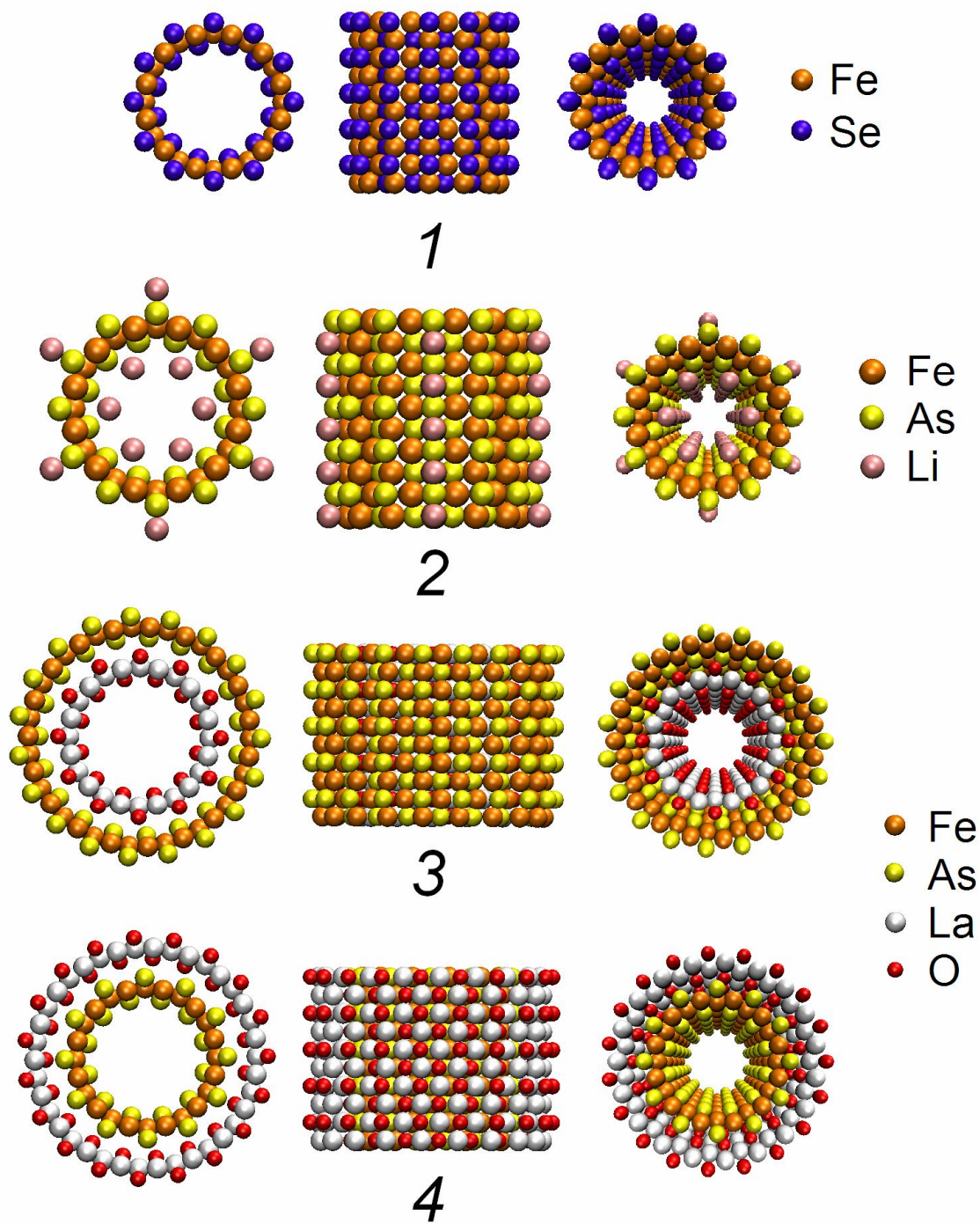

**Figure 2.** (Color online) Atomic models of *1*: FeSe NTs; *2*: LiFeAs NTs with configuration Li$_{1/2}$@[FeAs]@Li$_{1/2}$ (see text), and two main possible configurations of LiFeAsO NTs: arsenic-terminated (3) and oxygen-terminated (4), see text. For all of NTs their top view, side view and perspective view are shown.

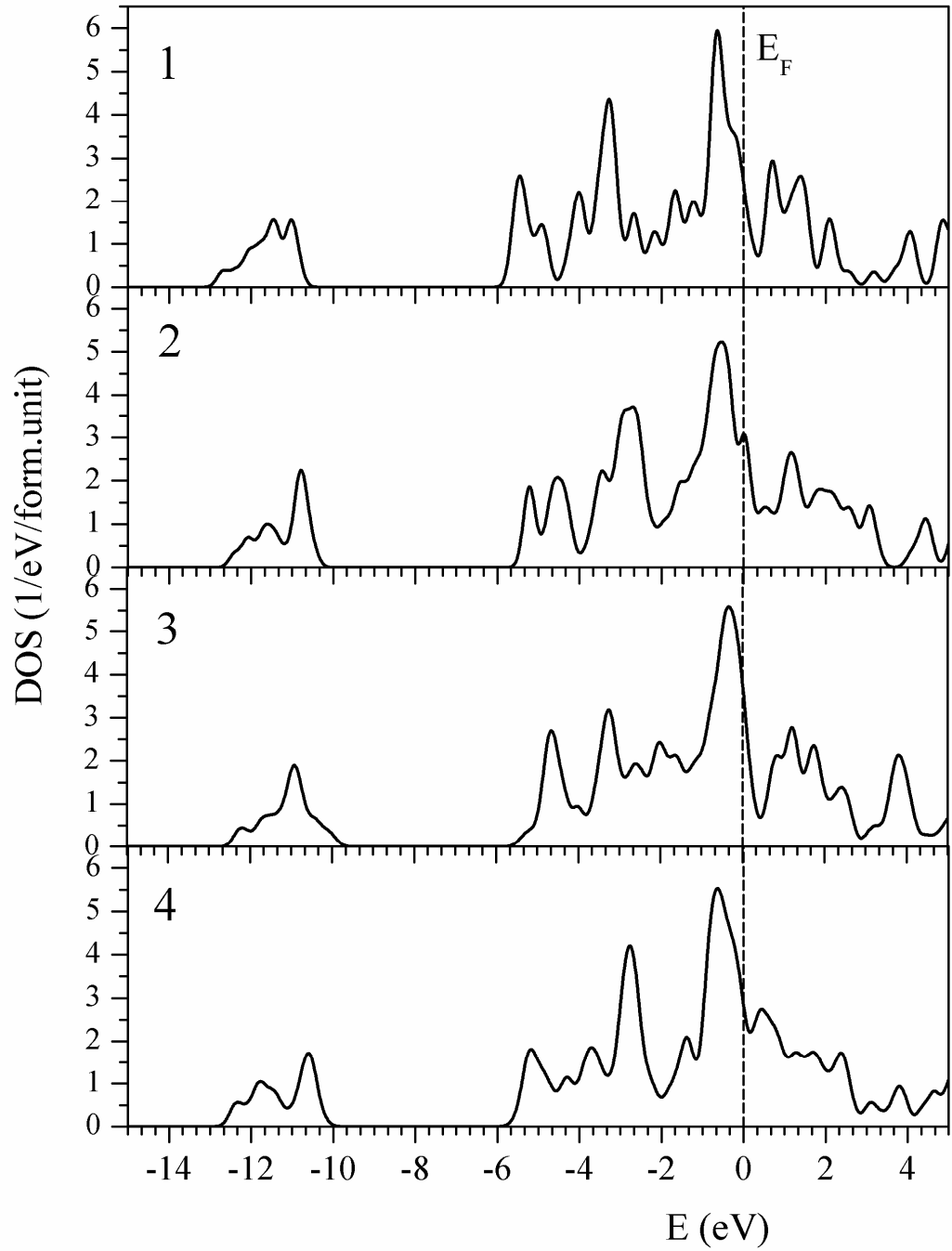

**Figure 3.** Total densities of states for the bulk $SrFe_2As_2$ (*1*), flat slab $SrFe_2As_2$ (*2*) and "*armchair*" (9,9) $SrFe_2As_2$ tubes with two main possible atomic configurations: 3 - [FeAs]@Sr, and 4- Sr@[FeAs], see also text.